# Characterization of Carbon-Contaminated $B_4C$-Coated Optics after Chemically Selective Cleaning with Low-Pressure RF Plasma

H. Moreno Fernández[a*], D. Rogler[b], G. Sauthier[c],
M. Thomasset[d], R. Dietsch[b], V. Carlino[e], E. Pellegrin[a*]

[a]CELLS-ALBA, Carrer de la Llum 2-26, E-08290 Cerdanyola del Valles, Spain
[b]AXO DRESDEN GmbH, D-01237 Dresden, Germany
[c]ICN2, UAB Campus, E-08193 Bellaterra, Spain
[d]SOLEIL Synchrotron, L'Orme des Merisiers, F-91192 Gif-sur-Yvette, France
[e]ibss Group Inc., Burlingame, CA 94010, USA

**ABSTRACT**

Boron carbide ($B_4C$) is one of the few materials that is expected to be mostly resilient with respect to the extremely high brilliance of the photon beam generated by free electron lasers (FELs) and is thus of considerable interest for optical applications in this field. However, as in the case of many other optics operated at modern light source facilities, $B_4C$-coated optics are subject to ubiquitous carbon contaminations. These contaminations represent a serious issue for the operation of high performance FEL beamlines due to severe reduction of photon flux, beam coherence, creation of destructive interference, and scattering losses. A variety of $B_4C$ cleaning technologies were developed at different laboratories with varying success. We present a study regarding the low-pressure RF plasma cleaning of carbon contaminated $B_4C$ test samples via inductively coupled $O_2$/Ar, $H_2$/Ar, and pure $O_2$ RF plasma produced following previous studies using the same IBSS GV10x downstream plasma source. Results regarding the chemistry, morphology as well as other aspects of the $B_4C$ optical coating before and after the plasma cleaning are reported. We conclude from these comparative plasma processes that pure $O_2$ feedstock plasma *only* exhibits the required chemical selectivity for maintaining the integrity of the $B_4C$ optical coating.

*Corresponding author. Tel.: +34 93 592 4418. E-mail: epellegrin@cells.es (Eric Pellegrin)

# 1. INTRODUCTION

Boron carbide ($B_4C$) as an engineering material has a long track record of applications in various fields of applications [1], although some of its polytypism is still not under full control, which is obviously due to the chemical similarity between its B and C atomic constituents. Nevertheless, optical engineering is presently increasing the usage of $B_4C$ as an optical coating material for a large range of optical application [2], including beamline optics in accelerator-based light sources such as synchrotron as well as free electron laser (FEL) facilities together with materials of similar hardness such as, e.g., SiC, cubic BN etc. where $B_4C$ is preferably used in the soft x-ray photon energy range (0.5 to 2.5 keV photon energy) while SiC is employed in the hard x-ray range (2 to 20 keV). This usage includes single coatings as well as multilayer mirror coatings based on, e.g., SiC/$B_4C$ multilayers.

While the cleaning of carbon contaminations of beamline optics based on metallic reflective coating has become a widespread activity [3–5](still occasionally suffering from heavy setbacks due to a lack of control and understanding of the basic cleaning process parameters), more complex optical surfaces and coatings such as, e.g., amorphous carbon, diamond-like carbon (DLC), silicon carbide (SiC), and boron carbide ($B_4C$) are now at the center of interest due to their unique capability of withstanding the high brilliance of the pulsed photon beam emitted from FEL light sources, which – at the same time – leads to higher carbon contamination rates on the optical surfaces within FEL beamline optics.

On the other hand, the extreme requirements imposed by FEL applications on the quality of beamline optical components and their preservation (especially when it comes to coherence-based experiments) imperatively call for the development of in-situ cleaning processes that warrant for the safe, efficient, and well-understood cleaning of FEL optical components with carbon-based optical coatings (e.g., such as the above $B_4C$, SiC etc.). Obviously, as carbon is present in both the coating as well as the surface contamination the cleaning technique to be used should necessarily include an inherent chemical selectivity for distinguishing these two different carbon species from each other in order to prevent any harm from the optical coating. To our present knowledge, several attempts have been performed so far regarding a cleaning of $B_4C$-coated optics using oxygen- as well as ozone-based in-situ and ex-situ techniques, but all of them so far did lead to a degraded $B_4C$ coating and were thus not deemed acceptable.

In this paper, we describe an experimental approach based on the low-pressure RF downstream plasma cleaning of various $B_4C$-coated test objects, which includes a variation of the plasma chemistry (and thus its chemical activity) by a variation of the plasma feedstock gases. Although the employed experimental procedure is similar to the one described in previous studies [6,7], the present results turn out to be completely different from those based on the cleaning of noble and non-noble metallic reflective optical coatings. In more detail, the results obtained give evidence for a satisfactory cleaning performance only for pure $O_2$ plasma, while an $O_2$/Ar feedstock gas mixture results into an efficient carbon cleaning as well, but combined with significant degradation of the $B_4C$ layer. The latter effect is even more pronounced in the case of Ar/$H_2$ feedstock gas mixtures, leading to a substantial reduction of the $B_4C$ layer thickness. We tentatively attribute these findings to a detrimental effect from the *kinetic* contribution from the plasma induced by especially the Ar species within the plasma. Thus, avoiding plasma species with elevated masses (such as, e.g., Ar) reduces the *kinetic* contribution within the plasma cleaning process, thus emphasizing the *chemical* contribution to the cleaning process while at the same time reducing its detrimental effect on the optical coating. In addition, we did find evidence for the plasma-induced surface conversion process of the $B_2O_3$ phase inherent to the $B_4C$ bulk layer into surface boron oxy-carbides, resulting into a significant reduction of cleaning rates.

## 2. EXPERIMENT

### 2.1 Coating processes

**$B_4C$ coating**

Amorphous $B_4C$-layers with a nominal thickness of 30±5 nm were fabricated by AXO Dresden GmbH on standard Si(100) wafers (10 x 10 mm$^2$ size) and optically polished single-crystalline Si substrates (<0.2 nm rms micro-roughness). All substrates were provided with a 2 nm (nominal thickness) Cr buffer layer between the Si substrate and the $B_4C$ coating. The $B_4C$ layer was deposited using a Dual Ion Beam Sputtering (DIBS) device from Roth & Rau AG (IONSYS1600). For the synthesis of the investigated $B_4C$-layers a sputter ion energy of 1000 eV was applied, which yields a $B_4C$ deposition rate of 0.2 Å/s.

**Amorphous carbon contamination coating**

For the deposition of amorphous carbon (a-C) layers – i.e., for simulating a photon-beam induced carbon contamination - onto the above $B_4C$ test samples a commercial e-beam deposition chamber has been used, that allows for a deposition of about 180 nm of carbon within about 200 seconds deposition time. Typical examples regarding the results from that amorphous carbon coating process in the specific case of the present $B_4C$ test samples can be seen in Figs. 1 and 2.

### 2.2 Si wafer and mirror test objects & systematic approach

**Analytical techniques**

The following techniques were used for investigating the $B_4C$-coated test samples: SEM, EDX, XPS, XRR, XPS, and interference microscopy.

**Si test wafers**

Si(100) wafer pieces, coated with an amorphous $B_4C$ layer (see Fig. 1) were used in order to characterize changes in the $B_4C$ surface chemistry as induced by the plasma treatment using x-ray photoemission spectroscopy (XPS). In addition, an energy-dispersive x-ray analysis (EDX) and/or scanning electron microscopy (SEM) analysis was performed on these Si test wafers in order to obtain information on the bulk versus surface chemical stoichiometry as well as changes in the surface morphology of the $B_4C$ coatings. A partial/central amorphous carbon coating spot was applied in order to investigate the influence of the direct exposure of the plasma onto the $B_4C$ coating (i.e., with or without the amorphous carbon top coating).

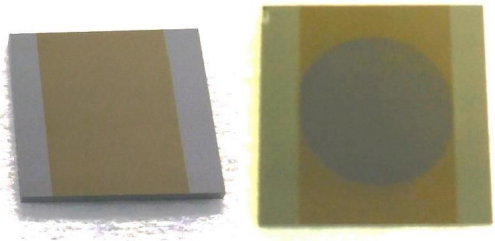

**Fig. 1:** Si(100) test wafers (10x10 $mm^2$ size) provided with a $B_4C$ coating stripe (ocher color – see left hand side panel) plus a central amorphous carbon contamination spot (dark grey color – see right hand side panel) on top of the $B_4C$ layer.

**Optical Test mirrors**

The second type of test objects consists of single-crystalline Si substrates with one inch diameter and an optical polishing with a surface micro-roughness of less than 0.2 nm rms. As in the case of the above Si(100) test wafers, the same $B_4C$ coating has been applied (see Fig. 2). Also, a partial amorphous carbon coating has been applied in order to investigate the influence of the direct exposure of the plasma onto the $B_4C$ coating. The surface micro-roughness measurements on these test mirrors were performed using a standard ZYGO interference microscope setup.

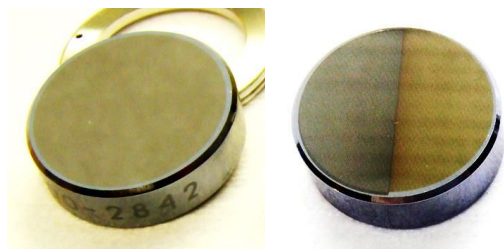

**Fig. 2:** Optically polished Si test mirrors with one inch diameter. The mirror on the left hand side shows a pristine $B_4C$ coating, while the mirror on the right hand side exhibits one half of its surface (i.e., the darker part) coated with an additional amorphous carbon layer with 70 nm thickness.

As in the previous studies [7,8] a standard quartz crystal microbalance (QCM) including gold-coated and a-C contaminated quartz crystals was used to determine the carbon removal rate during the plasma process for optimizing the plasma operation parameters towards the required optimum carbon cleaning rates.

### 2.3 Cleaning test setup

The test chamber setup for the plasma cleaning consists of a custom UHV chamber with a base pressure of $3 \times 10^{-7}$ mbar (unbaked), pumped by a combination of a turbo pump and a scroll pump in order to cover a wide range of gas pressures while actively pumping the system for cleaning studies utilizing a dynamic gas supply regime for the cleaning process. The complete cleaning chamber setup including the plasma source and plasma diagnostics are described in detail in a previous study [8].

### 2.4 GV10x inductively coupled RF plasma source

A commercial RF gun (model GV10x Downstream Asher, made by ibss Group, Inc., Burlingame, CA 940101, USA) has been used, based on the inductive coupling of the RF into the plasma tube inside the plasma source. This "downstream asher" RF plasma source runs a plasma in a separate volume upstream the chamber with the objects to be cleaned. This allows for an operation of the plasma chamber at higher pressures (plasma operation regime) with the cleaning chamber at lower pressures (i.e., larger reactant mean free path length) and, at the same time, reduces the detrimental kinetic effects from the plasma species in the downstream (or afterglow) plasma. In addition, the pressure gradient results in a gas jet of chemically active species between the two chambers volumes, thus facilitating the distribution of these active species across the downstream cleaning chamber volume.

The plasma parameters used for the different cleaning runs are given in Table I. Typical cleaning times required for an $O_2$/Ar or pure $O_2$ plasma were in the range of 2.5 hours, whereas the corresponding cleaning times for an Ar/$H_2$ plasma were in the range of 12.5 hours, starting from a-C coatings with the same thickness. This significant difference in terms of carbon cleaning rate between oxygen- and hydrogen-based downstream plasma by a factor of up to roughly seven could already be observed in previous studies [7,8].

| Gas Mixture | Working Pressure [mbar] | Feedstock Gas Ratio | RF Power [W] | Optical Emission Lines from Radicals | Emission Line Ratio |
|---|---|---|---|---|---|
| $O_2$/Ar | 0.005 | Ar 10% $O_2$ 90% | 100 | Ar I 750.4 nm | O* 95% Ar 5% |
| $H_2$/Ar | 0.002 | $H_2$ 65% Ar 35% | 100 | H I (H*) 656.3 nm | Ar 93% H* 7% |
| $O_2$ | 0.006 | $O_2$ 100% | 100 | O I (O*) 777.2 nm | -- |

**Table I:** Plasma parameters used for the different cleaning runs

### 2.5 $B_4C$ bulk thin film and surface characterization

The $B_4C$ test samples were analyzed by X-Ray Reflectometry (XRR) using Cu Kα radiation ($E_{Ph}$=8.04132 eV photon energy) before and after amorphous carbon coating plus subsequent RF plasma cleaning in order to detect changes in the $B_4C$ layer thickness and in the roughness of the $B_4C$/air interface. Simulations of the XRR results were perform using the IMD program [9]. Changes regarding the $B_4C$ sample surface chemistry were analyzed by XPS using a SPECS Phoibos 150 electron energy analyzer in conjunction with a monochromatized Al Kα x-ray source.

## 3. RESULTS AND DISCUSSION

### 3.1 $O_2$/Ar plasma cleaning

In Fig. S1 (see Supplementary Information file), we show $B_4C$-coated Si test wafers after the a-C deposition (180 nm thickness) as well after the subsequent cleaning with an $O_2$/Ar plasma. As can be seen from the visual appearance of the plasma-processed test wafer, a residue from the a-C spot can still be observed, that could not be fully removed without a visual degradation of the blank $B_4C$ coating. This is in stark contrast to previous findings on metallic optical coatings where a complete cleaning of the metal surface could be performed (i.e., without any visual residues) [8] and indicates the formation of a chemically stable that is more persistent the oxidative cleaning by the $O_2$/Ar plasma.

**X-ray Photoelectron Spectroscopy (XPS) analysis from $B_4C$-coated Si test wafer**

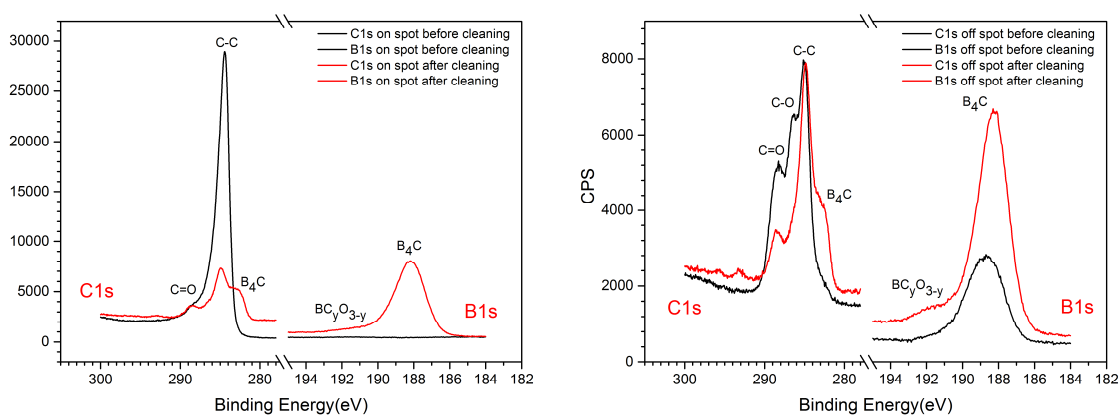

**Fig. 3:** High-resolution C1s and B1s XPS spectra of the $B_4C$-coated test wafer before (black solid lines) and after $O_2$/Ar RF plasma cleaning (red solid lines). Left panel: XPS data taken on the amorphous carbon contamination spot. Right panel: XPS data taken off the amorphous carbon contamination spot on the bare $B_4C$ coating.

Fig. 3 shows the XPS high resolution spectra from the Si test wafers in Fig. S1, with the measurements being performed off as well as on the a-C spot before and after the $O_2$/Ar plasma cleaning. We first consider the "on spot" spectra before/after the plasma cleaning. Here, one can clearly distinguish the absence and the appearance of the B1s XPS line before and after the plasma processing, respectively, together with a significant reduction of the C1s core level line.

The cleaning efficiency becomes clear from the increase (reduction) of the B1s (C1s) lines for the "on spot" via the cleaning process, as well as from the overall similarity of the B1s and C1s lines between the "on spot" and "off spot" locations after cleaning. Taking a closer look at the spectra, one can assign the different constituents of the XPS lines as shown in Table II. These assignments are in good agreement with the previous XPS study by Jacobsohn et al. and Jacques et al. [10,11].

| XPS line | C1s | | | | B1s | | | | O1s |
|---|---|---|---|---|---|---|---|---|---|
| B.E. [eV] | 282.58 | 284.94 | 286.6 | 288.58 | 188.12 | 189.49 | 191.40 | 193.3 | 533.0 |
| Assignment | C in $B_4C$ | C-C | C-O (e.g., in BOC) | C=O | B-B and B-C in $B_4C$ | B in $BC_2O$ | B in $BCO_2$ | B in $B_2O_3$ | O in $B_2O_3$ |

Table II: XPS core level line assignments ("BOC" refers to boron oxy-carbides)

According to the XPS line assignments given in Table II, one can interpret the high resolution spectra in such a way that the $O_2$/Ar plasma surface treatment leads to:

- A substantial reduction of the C-C C1s peak ("on spot" and "off spot") as well as the C-O and the C=O peaks ("off spot"), that are all related to adventitious or purposeful surface artifacts.
- The appearance of the B-B and B-C B1s line and the associated B-C peak in the C1s spectra ("off spot" and "on spot") being due to B and C in bulk $B_4C$, respectively.
- A weak B1s peak at 191.4 eV B.E. due to the occurrence of boron oxy-carbide (BOC).

On the other hand, we observe for the post-treatment spectra that the spectral fingerprint of especially the C1s XPS lines *invariably* exhibits the three peaks associated with $B_4C$, C-C, and C=O (with some minor shoulder at 286.6 eV due to C-O). This indicates the existence of a chemically stable layer at the $B_4C$ surface that is either resilient with respect to the plasma surface interaction and/or is a direct result from the latter. Similar results will be apparent from other plasma treatments with different feedstock gas combinations as will be shown in the next sections.

Using the same samples and systematics as used for the high resolution XPS spectra in Fig. 3, Fig. S4 shows the XPS survey spectra for the "on spot" and "off spot" sample locations before and after $O_2$/Ar plasma processing. Taking a look at the "off spot" survey spectra, one can distinguish an increase of the B1s/C1s line ratio together with the removal of adventitious N by the plasma treatment. Last but not least, the post-treatment "on spot" and "off spot" survey spectra display roughly the same intensities for the B1s, C1s, and O1s lines. A quantitative analysis of the $B_4C$-related B1s and C1s XPS lines shown in Fig. S8 gives a B/C atomic ratio of 4.19, which is close to the expected stoichiometry of 4. Summarizing, this gives evidence for an efficient cleaning of the a-C spot, without a significant oxidation of the $B_4C$ layer that would be evident from an increase of the O1s line.

Regarding the foreign materials other than those to be expected from $B_4C$, we note the occurrence of Au4f, N1s, and Fe2p lines. As the Au4f lines are present in the "off spot" survey spectrum, we attribute its occurrence due to an Au contamination during the $B_4C$ deposition process, while the N1s line is due to adventitious contaminations from exposure to the atmosphere being completely removed by the plasma process. In contrast to this, the Fe2p line - also being present in the "off spot" spectrum of the pristine material – is enhanced by the plasma process, thus raising the possibility of sputtering processes due to the heavy Ar species within the $O_2/Ar$ plasma.

**X-Ray Reflectometry (XRR) analysis**

In order to probe the effect of the $O_2/Ar$ plasma treatment onto the integrity of the $B_4C$ coating and the roughness of the $B_4C$/air interface, we have performed XRR measurements on both the Si wafer test coupons and the $B_4C$-coated test mirrors. The results from both types of test specimen are depicted in Figs. 4 and 5 and the parameters resulting from the IMD simulations of the experimental XRR data are given in Table III. All the XRR data simulation shown here include the Cr binding layer with about 2.2 nm thickness between the $B_4C$ coating and the Si substrate.

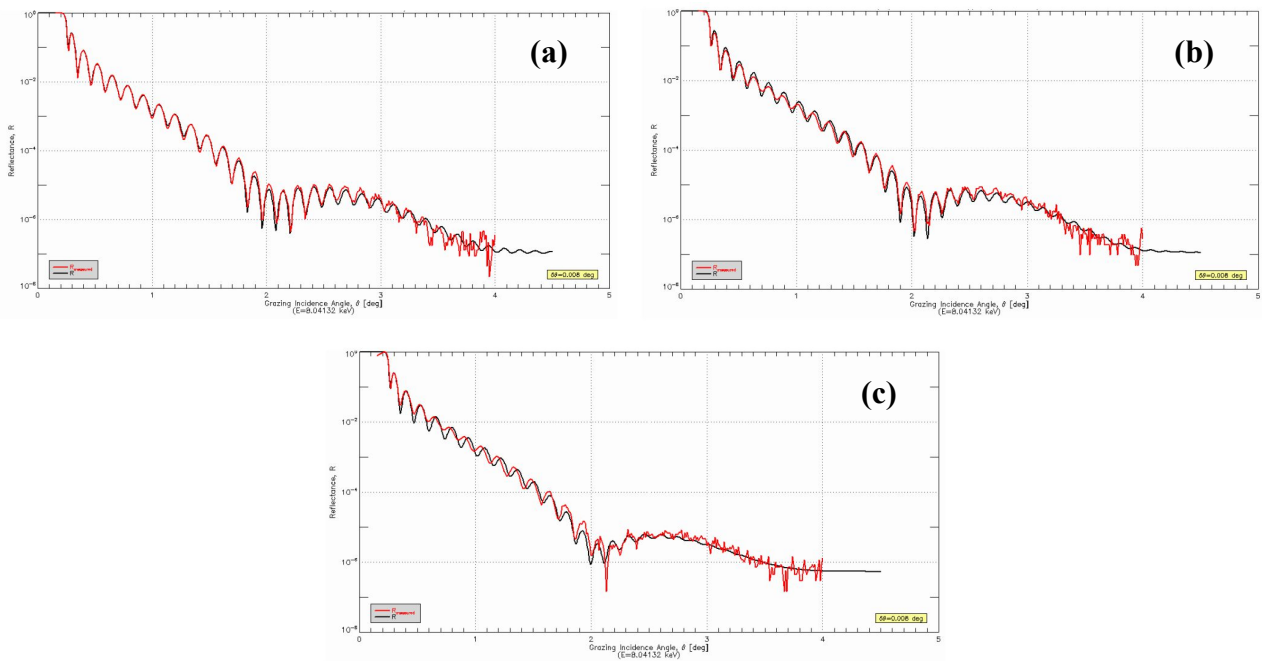

**Fig. 4:** XRR data from (a) pristine $B_4C$-coated Si test wafer right after $B_4C$ deposition, (b) after 7 months of storage in air, and (c) after $O_2/Ar$ plasma cleaning (red solid lines: experimental XRR data; black solid lines: IMD simulation)

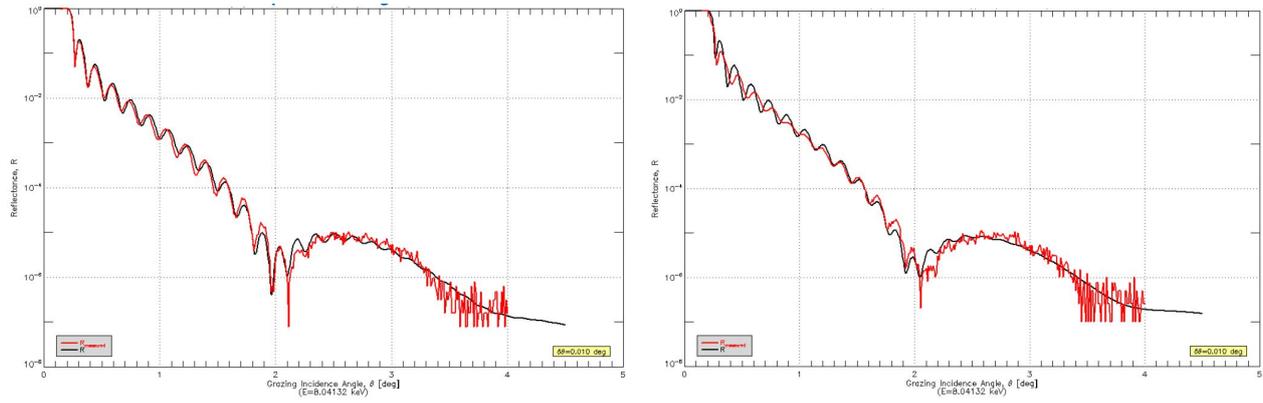

**Fig. 5:** XRR data from $B_4C$-coated test mirror right after $O_2$/Ar plasma cleaning. Left hand side: Non a-C coated part; right hand side: Formerly a-C coated part (red solid lines: experimental XRR data; black solid lines: IMD simulation)

| Si wafer test sample | $B_4C$ coating thickness [nm] | $B_4C$/air interface rms roughness [nm] | |
|---|---|---|---|
| After $B_4C$ deposition | **29.9** | ~ 0.4 | |
| After 7 months in air | 31.0 | ~ 0.5 | |
| After $O_2$/Ar plasma cleaning | **29.3** | ~ 0.6 | |
| **Mirror test sample** | $B_4C$ coating thickness [nm] | $B_4C$/air interface rms roughness [nm] | $B_4C$ rms surface roughness [nm] (*) |
| After $B_4C$ deposition | **26.0** | 0.5-0.6 | **0.11** |
| After 7 months in air | 26.9 | ~ 0.5 | 0.09 |
| After $O_2$/Ar plasma cleaning – non a-C coated part | 25.1 | ~ 0.6 | 0.12 |
| After $O_2$/Ar plasma cleaning – a-C coated part | **25.9** | ~ 0.7 | **0.17** |

**Table III:** Results from the IMD simulations of the XRR measurements as shown in Figs. 4 and 5. (*) Surface roughness values from interference microscopy.

When comparing the XRR results from the pristine Si test wafer (i.e., measured right after the $B_4C$ deposition) with those from the Si wafer stored in air for 7 months, there is an apparent increase in $B_4C$ layer thickness (i.e., 31.0 nm as compared to 29.9 nm). This is ascribed to the oxidation of the $B_4C$ surface due to the exposure to atmospheric air, leading to the formation of $BC_2O$ and $BCO_2$ together with the adsorption of adventitious carbon from atmospheric gases. Due to the limited chemical selectivity/sensitivity of the XRR analysis, we are left with the resulting apparent increase of the $B_4C$ layer by roughly 1.1 nm.

On the upside, the $O_2$/Ar plasma treatment of such a $B_4C$ layer after such an extended exposure to atmospheric air results in the removal of the physisorbed surface contaminations together with a reduction of the $B_4C$ layer thickness by 0.6 nm (i.e., 29.3 nm as compared to originally 29.9 nm). On the downside, the plasma treatment incurs an increase in rms surface roughness from 0.4 nm to 0.6 nm (see the upper part of Table III).

The XRR analysis from the $B_4C$-coated mirror in the lower part of Table III basically conveys the same message: A reduction of the $B_4C$ layer thickness between 0.9 and 0.1 nm (for the formerly C-coated and non C-coated part, respectively) together with an increase of rms surface roughness between 0.1 and 0.2 nm.

The obvious differences in terms of absolute surface roughness numbers between the results obtained from XRR and interference microscopy can be understood by the different level of sensitivity of these two techniques depending on the frequency range of the one-dimensional power spectral density function (PSD). The same argument applies when comparing, e.g., results from interference microscopy and AFM. Nevertheless, although being different in absolute size, both techniques should in principle observe the same trends.

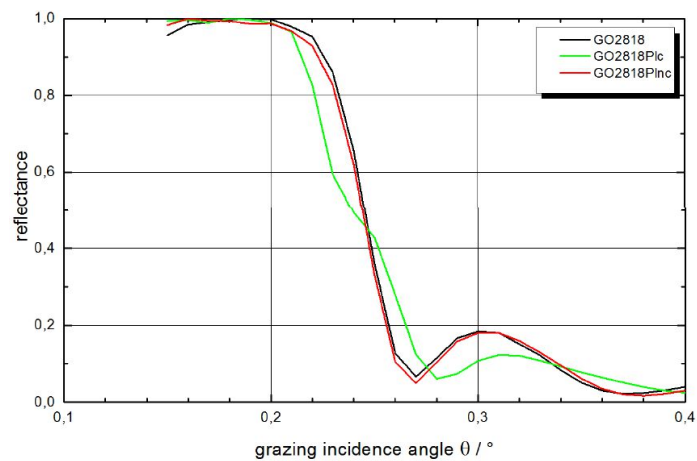

**Fig. 6:** Comparison of XRR total reflection data at low grazing angles for a pristine $B_4C$-coated test mirror (black line) and after a-C coating plus subsequent $O_2$/Ar plasma cleaning (green line: formerly a-C coated area; red line: non a-C coated area).

Additional information on the changes regarding especially the optical performance of the mirrors at Cu K$\alpha$ wavelength for correspondingly larger penetration depths (i.e., as compared to XPS) can be obtained by evaluating the total reflection part of the XRR data at low grazing incidence angles. In that respect, Fig. 6 gives evidence for an incomplete carbon cleaning of the mirror surface due to the deviation of the spectrum formerly a-C coated mirror surface (green line) at about 0.25 degree grazing incidence as compared to the spectra for the pristine and non a-C coated samples (black and red lines, respectively). Also, the further discrepancies around 0.3 degree grazing incidence corroborate the changes in $B_4C$ layer thickness mentioned above. The results from the surface roughness measurements via interference microscopy give – although being quantitatively lower than the XRR data – the same increase in terms of surface roughness by the plasma treatment (see Table III). Especially the deviations of the reflectivity edge of the formerly a-C coated material as compared to the pristine reflectivity are far from acceptable in terms of the expected optical performance.

**SEM results**

The $B_4C$ surface morphology of the Si test wafers in the SEM images shown in Fig. 7 give a visual account for the increase in rms surface roughness: Starting from the pristine test wafer in Fig. 7(a), one can distinguish small bright spots of about 10 nm diameter that are distributed across the surface in an irregular manner. Similar "hillocks" – but of micrometric size - have been observed in other studies [12] where they were attributed to crystalline $B_4C$ intergrowths in a B-doped pyrolytic carbon matrix formed during the post-growth heat treatment of $B_4C$ thin films as grown by CVD.

After a-C contamination and subsequent $O_2$/Ar plasma treatment of the Si test wafer, there is a clear increase regarding the density of these hillocks in the SEM image in the non a-C coated part (see Fig. 7(c)) that obviously leads to an increase of the surface roughness observed in the XRR data. Regarding Fig. 7(b), the residual carbon interface layer (see Fig. S1) apparently leads to lower density of these hillocks as compared to the formerly non a-C coated counterpart in Fig. 7(c). We attribute these hillocks to the formation of boron oxy-carbides that are actually more prominent in the case of $B_4C$ coatings treated with an Ar/$H_2$ plasma that will be discussed in a later section.

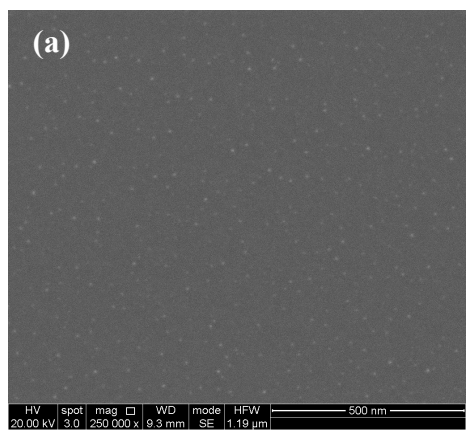

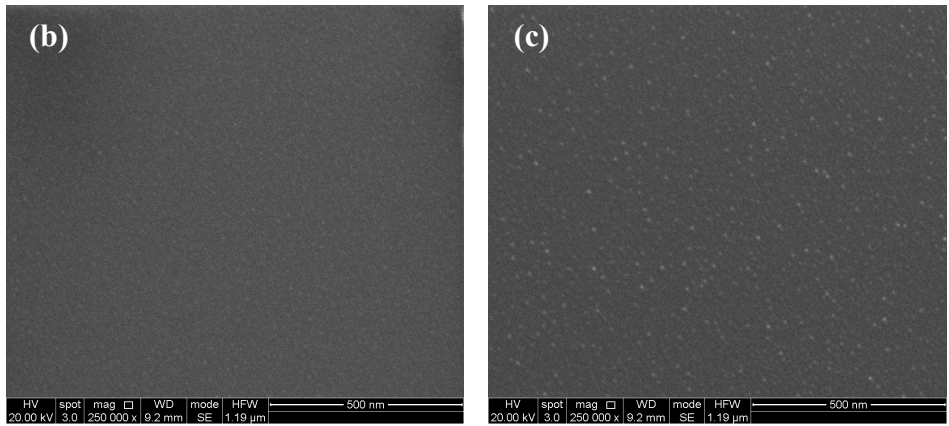

**Fig. 7:** SEM images of B$_4$C-coated test wafers taken at 20 kV electron acceleration voltage with a 250 000-fold magnification: (a) Pristine B$_4$C-coated test wafer, (b) O$_2$/Ar-plasma cleaned – formerly a-C coated part, and (c) O$_2$/Ar-plasma cleaned – non a-C coated part

Summarizing the results from the O$_2$/Ar plasma cleaning of B$_4$C-coated optical surfaces, this approach gives somewhat fair results in terms of preservation of the B$_4$C layer thickness and morphology, but the carbon cleaning appears to be somewhat poor as evidenced by the total refection XRR data as well as by the persistent non B$_4$C-related contribution in the C1s XPS spectra. This calls for an improvement of the present approach regarding the cleaning procedure.

### 3.2 H$_2$/Ar plasma cleaning

In this section, we follow the same approach as in the previous part, but this time using an H$_2$/Ar feedstock gas plasma that has been successfully used for the cleaning of optical surfaces coated with non-noble metals such as, e.g., Rh or Ni that are inherently incompatible with an oxidizing plasma as the latter would result in an oxidation of the optical coatings (thus merely replacing the carbon contamination with an oxide contamination).

In Fig. S2, we show the a-C coated Si wafer and mirror test items before (left hand side) and after (right hand side) the H$_2$/Ar plasma cleaning. According to the visual impression, the a-C residue appears to be less than in the case of the O$_2$/Ar-plasma cleaned test objects (see Fig. S1). However, especially in the case of the test mirror the non a-C coated part appears to have lost some of the ocher color from the original B$_4$C coating.

**X-ray Photoelectron Spectroscopy (XPS) analysis**

The high resolution XPS spectra in Fig. 8 yield the same results as their O$_2$/Ar analogues in Fig. 3: An almost complete removal of the a-C contamination, including a persistent three peak structure in the C1s XPS range. However, we note that in this specific case a significant increase of the B1s peak at 191.6 eV B.E. occurs that we relate to boron oxy-carbides.

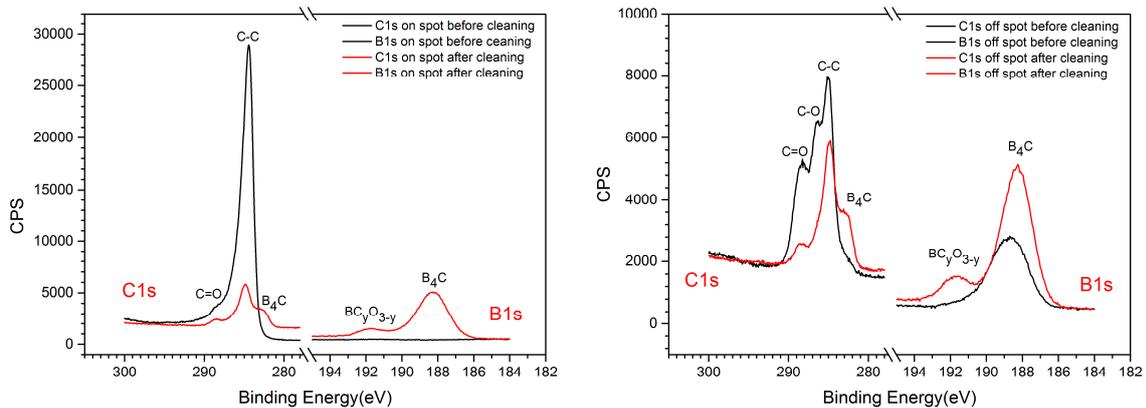

**Fig. 8:** High-resolution C1s and B1s XPS spectra of the B$_4$C-coated test wafer before (black solid lines) and after H$_2$/Ar RF plasma cleaning (red solid lines). Left panel: XPS data taken on the amorphous carbon contamination spot. Right panel: XPS data taken off the amorphous carbon contamination spot on the bare B$_4$C coating.

The XPS survey scans in Fig. S5 before/after $H_2/Ar$ plasma cleaning basically give the same results as in the case of the analogue spectra for the $O_2/Ar$ plasma treatment in Fig. S1 and also yield an efficient a-C removal from the sample surface. This resilience of the O1s line is surprising in a way as one would expect a decrease of the oxygen surface species due to the chemically reducing character of the $H_2/Ar$ plasma. However, a separate XPS depth profiling analysis (see Fig. S7) did reveal that the amorphous $B_4C$ layer contains a significant amount of oxygen, most probably as $BC_2O$, $BCO_2$, and $B_2O_3$ so that the simultaneously occurring removal of $B_4C$ surface layers (see below) leads to a persistent presence of oxygen species on the sample surface. As with the $O_2/Ar$ plasma appeared the Fe2p line, we attribute the latter to the plasma process, most probably due to sputtering phenomena related with heavy Ar species within the $H_2/Ar$ plasma.

**X-Ray Reflectometry (XRR) analysis**

The significant drawbacks from the $H_2/Ar$ plasma cleaning become apparent from the XRR measurements presented in Fig. 9. As is already obvious from the data, there is an almost complete loss of Kiessig fringes throughout the complete angular range, which applies to both the non a-C coated and the formerly a-C coated part. Taking a closer look at the results from the IMD simulations in Table IV, one can conclude that there has been a substantial reduction of the $B_4C$ layer thickness between about 8 and 4 nm for the non a-C coated and the formerly a-C coated part, respectively, together with a large increase of the $B_4C$/air interface roughness beyond 3 nm rms. These latter large numbers are probably partly due to some sample non-uniformities that may well contribute to the surface roughness. Nevertheless, the contribution from surface roughness itself is still substantial.

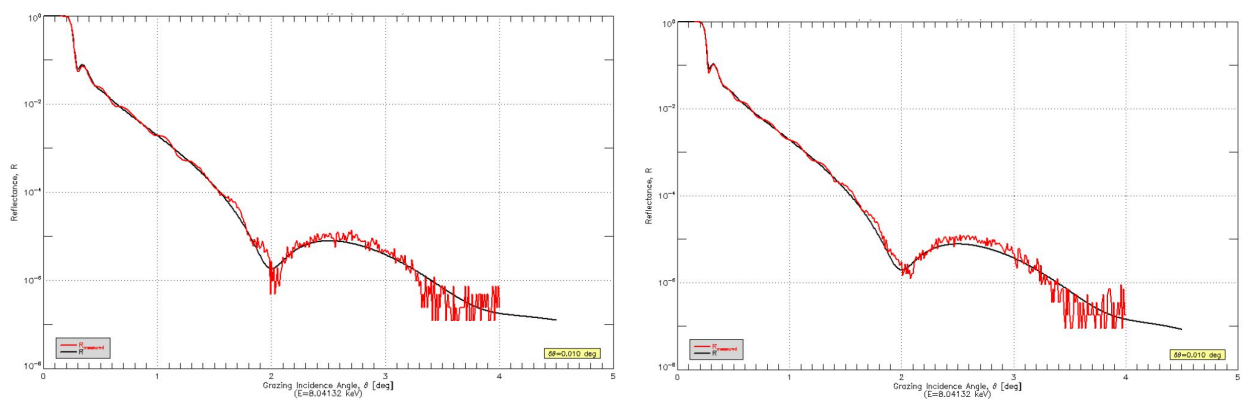

**Fig. 9:** XRR data from $B_4C$-coated test mirror right after $H_2/Ar$ plasma cleaning. Left hand side: Non a-C coated part; right hand side: Formerly a-C coated part (red solid lines: experimental XRR data; black solid lines: IMD simulation)

The surface micro-roughness data from interference microscopy (see Table IV) give a quantitatively more realistic picture, but still indicating a significant increase by the plasma treatment.

| Mirror test sample | B$_4$C coating thickness [nm] | B$_4$C/air interface rms roughness [nm] | B$_4$C rms surface roughness [nm] (*) |
|---|---|---|---|
| After B$_4$C deposition | 26.0 | ~ 0.5-0.6 | 0.12 |
| After H$_2$/Ar plasma cleaning – non a-C coated part | 18.2 | ~ 3.4 | 0.15 |
| After H$_2$/Ar plasma cleaning – a-C coated part | 21.7 | ~ 3.5 | 0.17 |

**Table IV:** Results from the IMD simulations of the XRR measurements as shown in Fig. 11.
(*) Surface roughness values from interference microscopy.

The total reflection data shown in Fig. 10 again indicate the structural changes for the B$_4$C layer at angles beyond 0.26 degree, while from the chemical perspective the deviation in the XRR data at about 0.25 degree indicate a satisfactory cleaning, which does not come as a surprise taking into account the loss of B$_4$C layer thickness mentioned above. Nevertheless, the clearly observable changes in the reflectivity edge for the plasma treated samples as compared to the pristine samples (i.e., distinct changes of the edges slope plus the occurrence of additional steps at about 0.18 degree grazing angle) indicate a severe and presumably unacceptable change of the optical performance of the B$_4$C coating due to, e.g., a chemical residue from the plasma cleaning, a change of the B$_4$C density, and/or B$_4$C surface roughness.

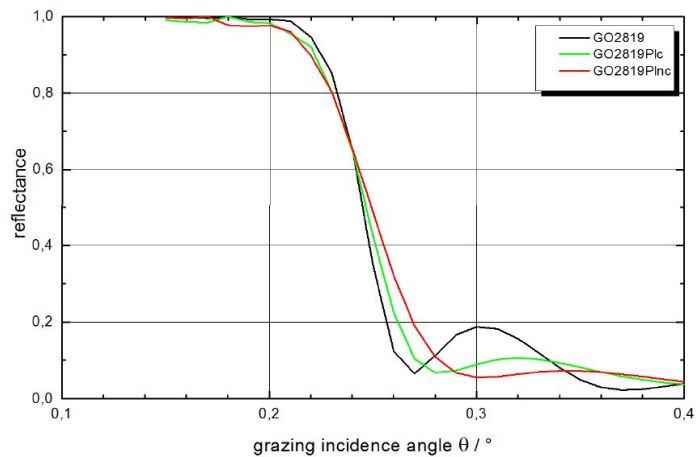

**Fig. 10:** Comparison of XRR total reflection data at low grazing angles for a pristine B$_4$C-coated test mirror (black line) and after a-C coating plus subsequent H$_2$/Ar plasma cleaning (green line: formerly a-C coated area; red line: non a-C coated area).

**SEM results**

The SEM results from a $H_2$/Ar -cleaned $B_4C$-coated test coupon are depicted in Fig. 11. Comparing these results with their analogues from the $O_2$/Ar plasma cleaning in Fig. 7, the images taken in the formerly a-C coated part appear to be very similar (i.e., visually indicating a carbonaceous residue on the samples surface), whereas the non a-C coated part of the $H_2$/Ar-cleaned $B_4C$-coated test coupon exhibits fewer, but more distinct/larger hillocks that we attribute to the boron oxycarbides that appears more strongly in the B1s XPS line in Fig. 8 ("off spot" after cleaning). It is obviously somewhat surprising that these hillocks can still be observed after a reducing $H_2$/Ar plasma process, but our XPS depth profile analysis shows that the $B_4C$ bulk material within the layers include quite some boron oxide phase that emerges to the surface during the sputtering by the Ar species within the plasma. Thus, we presume that there is a plasma-induced formation of boron oxy-carbide hillocks that are created by the interaction of the plasma with the $B_2O_3$ phases and $B_4C$ phases emerging at the sample surface.

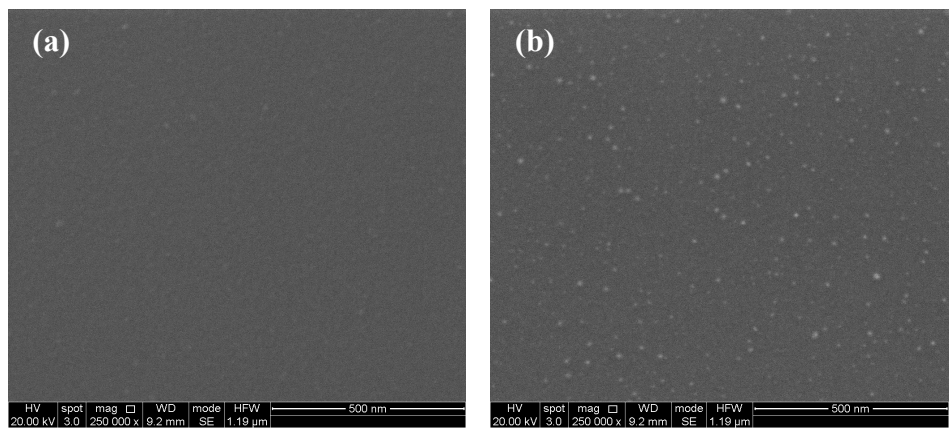

Fig. 11: SEM images of $B_4C$-coated test wafers taken at 20 kV electron acceleration voltage with a 250000-fold magnification after cleaning with $H_2$/Ar plasma: (a) Formerly a-C coated part, and (b) non a-C coated part

Summarizing the results from the $H_2$/Ar plasma cleaning of $B_4C$-coated optical surfaces, this approach gives contrary results as compared to those from the $O_2$/Ar plasma cleaning described in the previous section: Fair results in terms of carbon cleaning, but very bad results regarding the preservation of the $B_4C$ layer thickness and morphology as evidenced by the total refection XRR data. This also calls for an improvement of the present approach.

### 3.3 Pure $O_2$ plasma cleaning

In this section, we describe the results from the low-pressure RF plasma cleaning using pure $O_2$ feedstock gas, as the previous sections based on either $O_2$/Ar or $H_2$/Ar plasma cleaning did not really provide overall satisfactory results. In Fig. S3, we show a $B_4C$-coated Si test mirror during various stages of the experimental process. As can be seen from these photographs, the fully processed mirror compares visually quite well to the pristine mirror, which gives a first positive indication regarding the completeness of the cleaning process as the preservation of the $B_4C$ coating.

**X-ray Photoelectron Spectroscopy (XPS) analysis**

In Fig. 12, we show the high resolution C1s and B1s XPS spectra taken off as well as on the a-C spot before and after the pure $O_2$ plasma cleaning. As can be observed from the C1s spectra, there is a strong reduction of the C1s lines related to C=O, C-O, and C-C species for both the "off spot" and the "on spot" spectra due to the plasma cleaning process, whereas the B1s line to 188.3 eV B.E. related to $B_4C$ shows up after the plasma treatment. In the case of the B1s spectra, the "off spot" spectra exhibit a shift of the B1s line to lower B.E. (plus a decrease of the line width) and the appearance of a small additional peak at about 192.2 eV, that we attribute to oxidized boron and boron oxy-carbides, respectively.

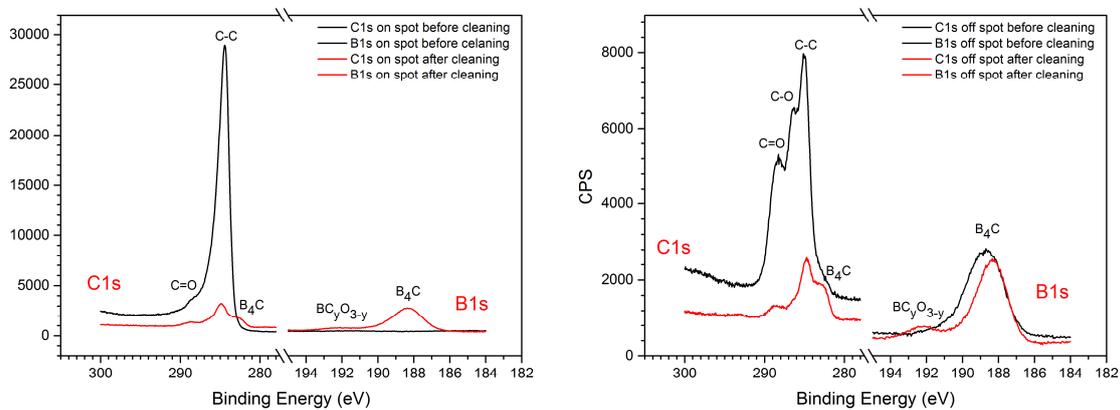

**Fig. 12:** High-resolution C1s and B1s XPS spectra of the $B_4C$-coated test wafer before (black solid lines) and after pure $O_2$ RF plasma cleaning (red solid lines). Left panel: XPS data taken on the amorphous carbon contamination spot. Right panel: XPS data taken off the amorphous carbon contamination spot on the bare $B_4C$ coating.

Fig. S6 shows the XPS survey spectra from $B_4C$-coated Si test wafers with the XPS measurements being performed off as well as on the a-C spot before and after the pure $O_2$ plasma cleaning. As observed in the previous sections, the "on spot" spectra give evidence for an efficient removal of the previously deposited a-C layer based on the significant reduction of the C1s line together with the appearance of the B1s line from the B$_4$C layer. The same applies to the "off spot" spectra, although with an obviously less spectacular reduction of the C1s line (as we are dealing with "off spot" spectra).

Regarding foreign elements, we again note the occurrence of a small amount of Au4f lines that have been identified as intrinsic to the B$_4$C coating as they are already present in the "off spot" spectrum of the sample before plasma treatment. It can also be noted that there is no Fe2p line visible in the survey spectra of the plasma-processed samples. This is in contrast to the previous section and can be attributed to the fact that there has been no Ar gas added to the pure O$_2$ feedstock gas mixture, which would otherwise lead to the aforementioned sputtering phenomena resulting in the deposition of a small amount of Fe.

**X-Ray Reflectometry (XRR) analysis**

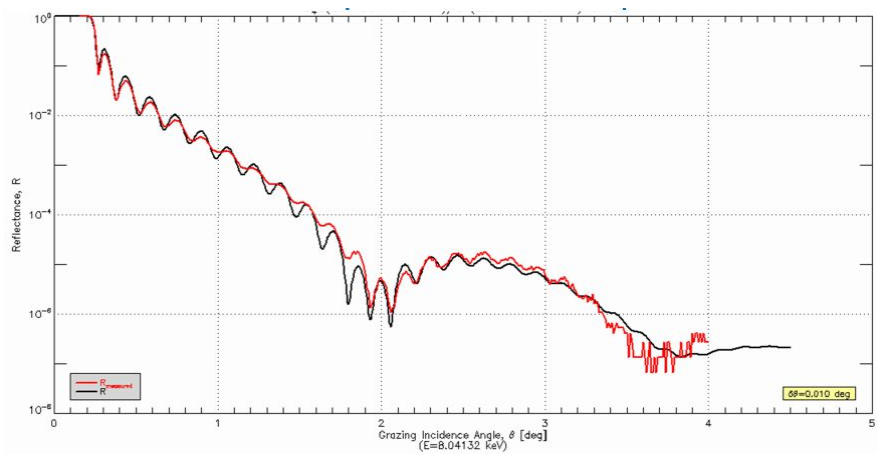

**Fig. 13:** XRR data from B$_4$C-coated test mirror after cleaning with pure O$_2$ plasma cleaning. (red solid lines: experimental XRR data; black solid lines: IMD simulation)

The measured XRR data in Fig. 13 together with the results from the IMD simulations given in Table V fully corroborate the results from the visual impression by the photographs in Fig. S3. As can be seen from the measured data, the Kiessig fringes throughout the full angular range are maintained and the parameters resulting from the pertinent IMD simulations yield a slight reduction of the B$_4$C coating thickness (i.e., 25.5 nm as compared to 26.0 nm) together with an unaltered B$_4$C/air interface roughness of 0.5 nm rms.

On the other hand, the micro-roughness data from interference microscopy show a substantial increase from 0.1o to 0.26 nm rms, which is at variance with the results from XRR. We attribute these contrasting trends to the technical differences between these two surface characterization techniques that will be discussed in the context of the SEM images.

| Mirror test sample | $B_4C$ coating thickness [nm] | $B_4C$/air interface rms roughness [nm] | $B_4C$ rms surface roughness [nm] (*) |
|---|---|---|---|
| After $B_4C$ deposition | **26.0** | ~ 0.5-0.6 | 0.10 |
| After 7 months in air | 27.0 | ~ 0.6 | -- |
| After pure $O_2$ plasma cleaning – a-C coated part | **25.5** | ~ 0.5 | 0.26 |

**Table V:** Results from the IMD simulations of the XRR measurements as shown in Fig. 16.
(*) Surface roughness values from interference microscopy.

The XRR total reflection data from a $B_4C$-coated test mirror in Fig. 14 give further support for a full recovery of the $B_4C$ surface at negligible losses regarding the integrity of the $B_4C$ layer. The spectrum for the a-C coated (i.e., non plasma-cleaned) mirror surface shows the expected chemical and structural deviations in the range from 0.2 to 0.26 degree and beyond 0.26 degree, respectively, due to the additional a-C top layer. On the other hand, the spectrum of the subsequently plasma cleaned sample nicely overlaps with the spectrum from the pristine untreated mirror in terms of the shape and slope of the reflectivity edge, beside a small shift regarding the edge position to slightly higher angular values. This leads to the conclusion for a full chemical and structural recovery of the test mirror from the perspective of Cu K$\alpha$ wavelengths and also applies to the optical performance of the fully processed mirror. Nevertheless, taking into account nowadays $B_4C$ optical coating manufacturing technologies a rms roughness figure of 0.26 nm appears acceptable.

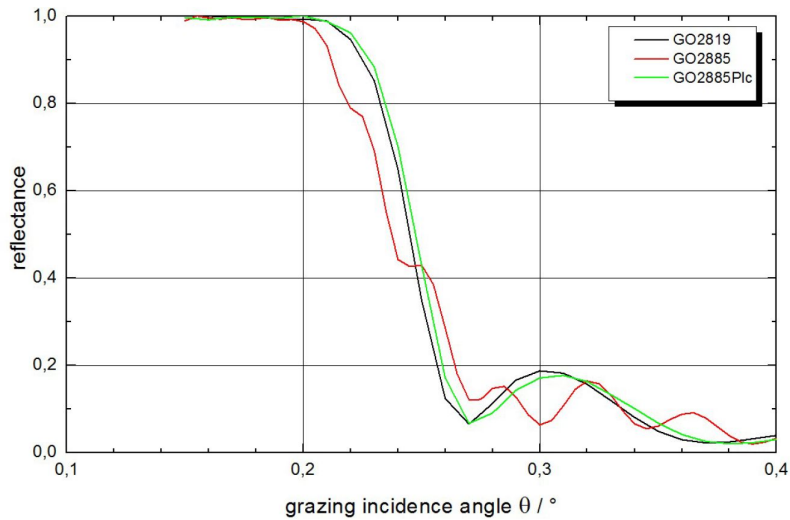

**Fig. 14:** Comparison of XRR total reflection data for a pristine $B_4C$-coated test mirror (black line), after a-C coating (red line), and after subsequent cleaning with pure $O_2$ plasma (green line).

**SEM results**

In Fig. 15, we show the SEM images of a fully processed $B_4C$-coated test coupon that has been cleaned with pure $O_2$ plasma. What can be immediately concluded from the images is that whereas the appearance of the non a-C coated part in Fig. 15(b) is similar to that of the $B_4C$ reference sample in Fig. 7(a) as well as the corresponding images from the other plasma processed samples in Figs. 7(c) and 11(b). In stark contrast to this, the image from the formerly a-C coated part in Fig. 15(a) exhibits a large density of small structures with diameters between 15 and 20 nm. As according the XPS results given in Fig. 12 the surface chemistry of the formerly a-C coated part is very similar to the ones from the $O_2$/Ar and $H_2$/Ar plasma treatment (see Figs. 3 and 8, respectively), we conclude that the surface modification in Fig. 15(a) is of a pure morphological nature, without apparently alternating the surface chemical characteristics that are basically given by the aforementioned plasma-induced conversion of the $B_4C$ and $B_2O_3$ bulk phases into boron oxy-carbides.

As mentioned above, the $B_4C$/air interface roughness as measured with XRR remains unchanged (see Table V), whereas the interference microscopy gives an increase from 0.10 to 0.26 nm rms. Taking into account the rather enhanced density of hillocks in Fig. 15(a) - that almost equals a full surface coverage - together with the fact that XRR is measured at grazing incidence, it seems plausible that interference microscopy (close to normal incidence) gives the more reliable surface roughness information, consistent with the SEM results.

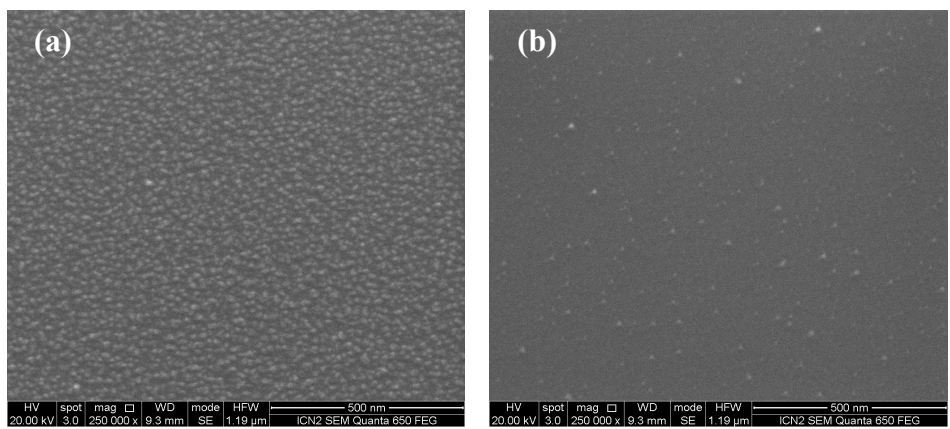

**Fig. 15:** SEM images of $B_4C$-coated test wafers taken at 20 kV electron acceleration voltage with a 250000-fold magnification after cleaning with pure $O_2$ plasma: (a) Formerly a-C coated part, and (b) non a-C coated part

Summarizing the results from the pure $O_2$ downstream plasma cleaning, we conclude that this approach is so far the only promising pathway that leads to an efficient cleaning together with the desirable preservation of the $B_4C$ surface morphology and chemistry, together with a conservation of the $B_4C$ bulk layer thickness. This also applies to the expected optical performance of the $B_4C$ coating as derived from the total reflection data. Nevertheless, the applied plasma cleaning time still has to be carefully taken into account in view of avoiding an over-cleaning of the surface coating.

### 3.4 Considerations on a-C plasma cleaning mechanisms and cleaning speed

In Table VI, we give a brief summary of the experimentally derived a-C cleaning rates that have been determined in the present study on a-C contaminated $B_4C$-coated test objects (extrapolated values) and on Au-coated QCM crystals that have been achieved in a previous study [6,7].

As has been observed previously, there is a large difference by roughly a factor of 7 between the a-C cleaning rates that can be achieved using either $O_2$/Ar or $H_2$/Ar feedstock gases on metallic substrates. One the other hand, it is clear from these numbers that while going from Au to $B_4C$ coatings, the cleaning rates for the $O_2$ and $O_2$/Ar-based plasma are strongly reduced (roughly by a factor of 2 to 2.5), whereas the cleaning rates for the $H_2$/Ar-based plasma is basically maintained.

| Plasma type and cleaning mechanism | $B_4C$ coated test coupons (this study; extrapolated) | Au-coated QCM crystal (previous study*) |
|---|---|---|
| Pure $O_2$ (chemical cleaning) | 3.5 Å/min. | 8.4 Å/min. |
| $O_2$/Ar (mostly chemical cleaning) | 4.6 Å/min. | 11.6 Å/min. |
| $H_2$/Ar (mostly kinetic cleaning) | 2.3 Å/min. | 1.7 Å/min. |

**Table VI:** Experimentally derived a-C cleaning rates for different types of plasma feedstock gases for either $B_4C$-coated test coupons and Au-coated QCM crystals. All values were measured using 100 W plasma RF power at a plasma vacuum pressure of 0.005 mbar (* = data taken from Ref. [6,7])

We basically attribute these changes in a-C cleaning rates for the different substrate coatings – i.e., more specifically the reduction of cleaning rates for the $O_2$-gas based plasma when going from Au to $B_4C$ coatings – to two different phenomena:

- The plasma-induced formation of a chemically resilient carbonaceous layer on the sample surface. This is corroborated by the observation of an ubiquitous non-removable and optically visible residue at the location of the formation a-C contamination as well as the resilience and, at the same time, the close similarity of the XPS spectra after the cleaning process.

- The fundamentally different processes that lead to the removal of the a-C contamination as a function of the chemical composition of the plasma feedstock gas. In the oxygen-rich/argon-poor $O_2$ and $O_2$/Ar plasma, we assign most of the cleaning to the chemical activity of the oxygen species within the plasma. On the other hand, in the case of the argon-rich $H_2$/Ar plasma, a significant (or may be even dominant) part of the a-C removal is done via the kinetic effect from the Ar species in the plasma as well as UV photochemical contributions from an Ar metastable state.

The above two phenomena apparently result in the reduction of carbon cleaning rates due to the chemical resilience of the carbonaceous BOC layer, especially in the case of the chemical cleaning using the oxygen-based plasma. In contrast to this, the cleaning speed by the argon-rich $H_2$/Ar plasma is maintained as the kinetic cleaning is kept up also in the case of the carbon-contaminated $B_4C$-coated optics, which is corroborated by our findings regarding the reduction of the $B_4C$ layer thickness due to the $H_2$/Ar plasma as discussed in section 3.2. The latter does obviously not necessarily exclude the reductive effect of the H˙ radicals that could be observed for the cleaning a-C contaminated metal foils, where a reduction of $Ni_2O_3$ and $Rh_2O_3$ to metallic Ni and Rh could be observed.

# 4. SUMMARY AND OUTLOOK

In this paper, we report on different approaches regarding the *chemically selective* low pressure RF downstream plasma cleaning of $B_4C$-coated and amorphous carbon-contaminated optics by the variation of the plasma feedstock gas and thus the inherent plasma chemistry. We thereby also report on the first successful plasma cleaning of such optics, i.e., without incurring damage of the $B_4C$ optical coating and its surface resulting into a fully maintained optical performance.

In more detail, we have performed an extensive test series on various test objects using $O_2$/Ar, $H_2$/Ar and pure $O_2$ gas mixtures as plasma feedstock gases in order to determine the optimum plasma chemistry for the cleaning of $B_4C$ optical surface coatings. The analysis of these $B_4C$ coatings and surfaces as a function of the a-C contamination and cleaning process has been performed using XPS, XRR, SEM, and interference microscopy in order to track changes in the $B_4C$ bulk as well as surface.

Based on our analysis, we conclude that the $O_2$/Ar plasma cleaning yields fair results in terms of the preservation of the $B_4C$ layer, while the removal of amorphous carbon contaminations can be considered as poor. On the other hand, the $H_2$/Ar plasma treatment leads to a significant damage of the $B_4C$ coating – both regarding the layer thickness and surface roughness – while the removal of amorphous carbon contaminations seems to be fairly satisfactory. However, it has to be taken into account that the latter is apparently an obvious consequence from the former. Last but not least, the plasma cleaning using pure $O_2$ feedstock gas yields excellent results regarding both the a-C cleaning as well as the preservation of the $B_4C$ layer.

From the above observations, we conclude that the Ar species within the plasma – especially in conjunction with the reduced a-C cleaning rates and thus enhanced cleaning times of an $H_2$/Ar plasma – result into sputtering phenomena on the $B_4C$ layers that lead to the observed reduction in layer thickness and enhanced surface roughness. Thus, a pure $O_2$ feedstock gas plasma (i.e., without admixture of heavy Ar gas atoms) offers a damage-free and fast cleaning, based on the chemical activity of the O species. First measurements using plasma mass spectroscopy indicate that the predominant oxygen species within an $O_2$/Ar feedstock gas plasma consist of $O_2^+$ species.

Almost independent from the plasma used, we have observed the formation of a fair amount of boron oxy-carbides on the $B_4C$ surfaces that presumably appear as hillocks on these surfaces and thus contribute to the surface micro-roughness. This especially applies to the $H_2$/Ar plasma-cleaned test samples. The formation of this chemically rather resilient carbonaceous layer could be observed both visually as well as based on XPS data and appears to contribute to the required extended cleaning times as compared to, e.g., the cleaning of metallic optical coatings. Everything considered, we conjecture that the formation of this surface boron oxy-carbide layer is the result from the surface interaction of the plasma with the $B_2O_3$ and $B_4C$ phases inherent to the bulk of the $B_4C$ coating, resulting into the formation of the said surface boron oxy-carbides.

The apparent sensitivity of the $B_4C$ layers with respect to a plasma-induced damage (i.e., as compared to apparently less sensitive metallic optical coatings) emphasizes the importance of an in situ monitoring of the cleaning process in view of avoiding a cleaning process beyond the time required for the removal of the carbon contaminations. To this end, an optical in situ end point detection scheme is presently under development that allows monitoring the progress of the cleaning process in order to stop the treatment once a clean optical surface has been achieved.

**Acknowledgements**

We acknowledge initial motivation and support of the $B_4C$ plasma cleaning project by R. Barrett (ESRF), the support by technical staff of the CELLS Synchrotron Light Source facility, especially the expert assistance by L. Ginés and D. Calderon (CELLS). The skillful technical assistance by M. Rosado (ICN2) for the SEM and EDX analysis is also gratefully acknowledged.

**Funding**

The research by HMF is supported by funding from the "Generalitat de Catalunya, Departament d'Empresa i Coneixement" within the "Doctorats Industrials" program (dossier no. 2014 DI 037).